\documentclass[10pt,twocolumn,letterpaper]{article}

\usepackage{cvpr}
\usepackage{times}
\usepackage{epsfig}
\usepackage{graphicx}
\usepackage{amsmath}
\usepackage{amssymb}

\usepackage{tabularx}
\usepackage{multirow}
\usepackage{hhline}
\usepackage{graphicx}

\usepackage[pagebackref=true,breaklinks=true,letterpaper=true,colorlinks,bookmarks=false]{hyperref}

\cvprfinalcopy % *** Uncomment this line for the final submission

\def\cvprPaperID{7} % *** Enter the CVPR Paper ID here

\ifcvprfinal\fi
\begin{document}

%%%%%%%%% TITLE
\title{Application of DenseNet in Camera Model Identification and Post-processing Detection}

\author{Abdul Muntakim Rafi\textsuperscript{1} \qquad Uday Kamal\textsuperscript{1} \qquad Rakibul Hoque\textsuperscript{1} \qquad Abid Abrar\textsuperscript{1} \qquad Sowmitra Das\textsuperscript{1} \qquad \\Robert Laganière\textsuperscript{2} \qquad Md. Kamrul Hasan\textsuperscript{1*}\\
\textsuperscript{1}Bangladesh University of Engineering \& Technology \qquad \qquad \textsuperscript{2}University of Ottawa\\
{\tt\small \textsuperscript{2}laganier@uottawa.ca \qquad \qquad \textsuperscript{1*}khasan@eee.buet.ac.bd} 
%\and
%{\tt\small xxxxxxxxxxxxx@xxx.com}
%\and
%Third Author\\
%Institution3\\
%{\tt\small xxxxxxxxxxxxx@xxx.com}
%\and
%Fourth Author\\
%Institution4\\
%{\tt\small xxxxxxxxxxxxx@xxx.com}
%\and
%Fifth Author\\
%Institution5\\
%{\tt\small xxxxxxxxxxxxx@xxx.com}
%\and
%Sixth Author\\
%Institution6\\
%{\tt\small xxxxxxxxxxxxx@xxx.com}
}

\maketitle
%\hypersetup{%
%	citecolor=black
%}
%\thispagestyle{empty}

%%%%%%%%% ABSTRACT
\begin{abstract}
	
	Camera model identification has earned paramount importance in the field of image forensics with an upsurge of digitally altered images which are constantly being shared through websites, media, and social applications. But, the task of identification becomes quite challenging if metadata are absent from the image and/or if the image has been post-processed. In this paper, we present a DenseNet pipeline to solve the problem of identifying the source camera-model of an image. Our approach is to extract patches of  $256\times 256$ from a labeled image dataset and apply augmentations, i.e., Empirical Mode Decomposition (EMD). We use this extended dataset to train a Neural Network with the DenseNet-201 architecture. We concatenate the output features for 3 different sizes ($64\times 64, 128\times 128, 256\times 256$) and pass them to a secondary network to make the final prediction. This strategy proves to be very robust for identifying the source camera model, even when the original image is post-processed. Our model has been trained and tested on the Forensic Camera-Model Identification Dataset provided for the IEEE Signal Processing (SP) Cup 2018. During testing we achieved an overall accuracy of 98.37\%, which is the current state-of-the-art on this dataset using a single model. We used transfer learning and tested our model on the Dresden Database for Camera Model Identification, with an overall test accuracy of over 99\% for 19 models. In addition, we demonstrate that the proposed pipeline is suitable for other image-forensic classification tasks, such as, detecting the type of post-processing applied to an image with an accuracy of 96.66\% -- which indicates the generality of our approach. 
\end{abstract}

%%%%%%%%% BODY TEXT
\section{Introduction}
In digital forensics, camera model identification is a distinguished field of research and has profound impact on crucial real-life applications, such as criminal investigations,  authenticating evidence, detecting forgery, etc. Nowadays, professional image editing tools are readily available, making image-forgery quite commonplace. Thus, cyber-crimes via digital images are ever increasing and; so as, the need for a robust camera model identification scheme. But unfortunately, the task of identifying the camera-model is very challenging, especially when the metadata of the digital image is unavailable. As a result, a forensic analyst has to implement unique techniques to determine the source camera-model solely from an image. \vspace{0.1mm}

In the literature, various methods have been proposed to perform this task. \cite{stamm2013information}, \cite{kirchner2015forensic} and \cite{piva2013overview} have perfectly described the present condition of camera model identification in their review. The initial approach was an infeasible idea of merging external features in an image for each and every device; like watermarks, device-specific-code, etc. As a result, focus has shifted towards detecting intrinsic camera features, such as the Color Filter Array (CFA) pattern (\cite{bayram2005source}), interpolation algorithms and Image Quality Metrics (IQM) used in the camera (\cite{kharrazi2004blind}, \cite{gloe2012feature}). Device-specific camera-detection schemes have also been proposed, where noise-patterns like the Photo Response Non-Uniformity (PRNU) have been exploited to identify the device (\cite{dirik2007source}, \cite{lukavs2006digital}), \cite{filler2008using}.  Although device specificity is an inherent feature of PRNU noise, forensic researchers have developed methods to make camera model identification device invariant (\cite{thai2014camera}, \cite{lukas2006digital}). Most of these works try to estimate the model-specific artifacts that are introduced into an image during image-capture, and then, correlate these features with a reference for the corresponding camera-model (\cite{cao2009accurate}). In this approach, the second order statistics of the CFA pattern (\cite{swaminathan2007nonintrusive}) and 3D co-occurrence matrices (\cite{chen2015camera}, \cite{marra2017study}) have been used as feature vectors to successfully detect camera-models with state-of-the-art accuracy. \vspace{0.1mm}

Most of the methods stated so far have used traditional complex ensemble classifiers. Recently, researchers have adopted a data-driven approach and made an effort to solve this problem using Convolutional Neural Networks (CNN). This suggestion seems quite promising because, in the past decade, Neural Networks have achieved phenomenal accuracy on image-classification benchmarks (\cite{schmidhuber2015deep}). To this end, \cite{tuama2016camera} have trained a CNN on the Dresden database to solve this classification problem. Their work also includes the use of preprocessing using a custom built 2D high-pass filter.  However, their overall accuracy is below the state-of-the-art accuracy reported in (\cite{chen2015camera}). A concept of Content Adaptive Fusion Network is introduced by \cite{yang2017source}, which is basically a cluster of CNNs with different kernel size, has been introduced to classify camera brand and device to achieve a moderate accuracy around 95\%. In spite of the breadth of work performed in this field, little attention has been given to the detection of camera-model specifically from post-processed images (such as different JPEG Compression Rate, Resized, Gamma-Corrected images etc.) Though researchers have explored some of these cases discretely, not many have tried to bring them into the same framework. Image authentication from JPEG headers (\cite{kee2011digital}), forgery detection from intrinsic statistical fingerprints of images (\cite{stamm2010forensic}), detecting doubly compressed JPEG images using Discrete Cosine Transform (DCT) (\cite{li2008detecting}), and even the recent use of CNN to detect image manipulation in (\cite{bayar2016deep} and \cite{bayar2017design}) are some examples of work done in detecting image manipulations and classifying them using this approach.  \vspace{0.1mm}

But still, the use of very deep networks is yet to be explored thoroughly for this task. In the absence of metadata and the presence of extensive post-processing in images, we believe that Deep Neural Networks (DNNs) have the potential to achieve a better classification-rate than existing methods. In the presence of these challenges, traditional feature-vectors such as the DCT-Residue(\cite{roy2017camera}) and co-occurrence matrices((\cite{chen2015camera}, \cite{marra2017study})) are unarguably altered, often in ways that cannot be predicted in the general case. Thus, designing features that retain the camera-model information even from post-processed images is quite cumbersome, if not, extremely difficult (\cite{lukas2006digital}). This provides the motivation to use Neural Networks to perform this task. Since, DNNs do not require explicit feature engineering, and can automatically learn the necessary features from the image, it makes the task of classification more tractable. \vspace{0.1mm}

As stated earlier, forensic researchers have used a number of custom neural network architectures (\cite{yang2017source}, \cite{bayar2017design}, \cite{bondi2017first}, \cite{yao2018robust}). Besides, a number of deep architectures have been also proposed to perform the task of classification, such as the VGGNet, GoogLeNet, ResNet and most recently, the DenseNet (\cite{marra2018vulnerability}). A major challenge in using such deep networks is to address the issues of over-fitting and feature-attenuation during training. The camera-model features existing within an image are extremely subtle, compared to other dominating features of the image. As such, while training a Deep Neural Network, these model-level features may be sharply attenuated as the input image is propagated through successive layers.  \vspace{0.1mm}

In this work, we choose to use the 201-layer DenseNet \cite{huang2017densely} as the core architecture of our network. In the DenseNet, the output of a certain layer is propagated to all the layers in front of it. Any layer in the network has direct access to the features generated by all the layers that came before it. As a result, if any of the image features are lost during forward propagation, they are re-generated at the input of latter layers through the dense connections. That is why, this architecture is quite suitable for detecting minute statistical features like those related to the source camera-model of the image. Again, experiments on image-classification benchmarks have shown that, using a secondary network to re-calibrate the learned features improves the representational power of a network. Motivated by these results, we feed the output features of our main network into a Squeeze-and-Excite block, introduced in (\cite{hu2017squeeze}). This boosts the final test-accuracy of our networks on the given benchmarks. \vspace{0.1mm}

But, the problem of over-fitting still remains, as existing datasets are limited in their size. To overcome this setback and  to ensure generalization of the features that are learned, we use a number of data-augmentation schemes such as, gamma correction, JPEG compression, re-scaling, extracting patches for training, randomly cropping and flipping the training image to extend the dataset. These prevent the network from becoming dependent on the specific device from which the training images are taken and help the network learn more robust features. Additionally, we have also trained the network for image manipulation detection, to see whether or not- it can be used as a general purpose network for image forensics.\\
The following sections of this paper explain image acquisition model, the outline of our model, detailed architecture of the network, training procedure, and the detailed comparative results.

\section{Materials}

\subsection{Description of Datasets}
In order to train our network, we have used the Camera-Model Identification Dataset provided for the IEEE Signal Processing (SP) Cup 2018. The initial dataset consisted of images captured by 10 different camera models having 275 images for each, all of which were provided by the IEEE Signal Processing Society. In addition to this, external data for each camera-model is collected from Flickr during the open competition phase of SP Cup 2018. This one contained varying number of images for each camera model. Dataset-I is formed by combining both of these sets of data. A brief summary of the dataset-I is given in Table \ref{datasum1}. 

\begin{table}[!ht]
	\centering
	\begin{tabular}{|r|c|c|}
		\hline
		Camera Model		&\multicolumn{1}{c|}{\begin{tabular}[c]{@{}c@{}c@{}}SP Cup Data\\ (No. of\\ Images)\end{tabular}}  	&\multicolumn{1}{c|}{\begin{tabular}[c]{@{}c@{}c@{}}Flickr Data\\ (No. of\\ Images)\end{tabular}} \\
		\hhline{|=|=|=|}
		HTC-1-M7			& \multirow{10}{1.3cm}{$\;\;$ 275\\ $\times$ 10}	&746 \\
		\cline{1-1}\cline{3-3} 
		iPhone-4s 			& 	&499\\
		\cline{1-1}\cline{3-3} 
		iPhone-6 			& 	&548\\
		\cline{1-1}\cline{3-3} 
		LG-Nexus-5x 		& 	&405\\
		\cline{1-1}\cline{3-3} 
		Motorola-Droid-Maxx & 	&549\\
		\cline{1-1}\cline{3-3} 
		Motorola-Nexus-6 	& 	&650\\
		\cline{1-1}\cline{3-3} 
		Motorola-X 			& 	&344\\
		\cline{1-1}\cline{3-3} 
		Samsung-Galaxy-Note3& 	&274\\
		\cline{1-1}\cline{3-3} 
		Samsung-Galaxy-S4 	& 	&1137\\
		\cline{1-1}\cline{3-3} 
		Sony-NEX-7 			& 	&557\\
		\hhline{|=|=|=|}
		Sub-Total 			& 2750 	&5709\\
		\hline
		Grand-Total			& \multicolumn{2}{c|}{8459}\\
		\hline
		
		%\multirow{2}{*}{Challenge Types} & 12\\
		%& + 1 (No Challenge)\\
		%\hline
	\end{tabular}
	\caption{SP Cup data \& Flickr data (Dataset I)}
	\label{datasum1}
\end{table}

The test data for the SP Cup dataset is provided separately on the Kaggle platform without any labels. It includes 2640 images of size $512\times 512$, among which 1320 are unaltered and the rest are manipulated externally. The details of the manipulation scheme used to generate these images are discussed in subsequent sections.

In addition to Dataset-I, we have also performed experiments on the  Dresden Image Database. This dataset include varying number of images for 27 different camera models. We denote these images as Dataset-II. 

\subsection{Data Augmentation}
Additional data has been generated by post-processing the original images given in the dataset. 
%Post-processing the images may seem like an added layer of challenge at first, but surprisingly, it turns out to be beneficial in the context of a Neural Network framework. 
It is a common practice in deep learning to deliberately alter the input data to help the network learn more robust features. 
%As a result, in this task, the Neural Network gets an added impetus due to the presence of the post-processed images, and the training procedure is further generalized.
A total of 8 types of post-processing have been performed on the images of Dataset-I. These are JPEG-Compression with quality factor 90\% and 70\%; Resizing by a factor of 0.5, 0.8, 1.5 and 2.0; Gamma-Correction using $\gamma$ as 0.8 and 1.2. Also, EMD has been performed as an augmentation which is discussed in section 2.2.2. Moreover, the input image is randomly rotated by $0^{\circ} ,\pm 90^{\circ}$, and $180^{\circ}$ during training. Because of this, the network can extract the camera model-features irrespective of whether the image was taken in landscape mode or portrait mode. 
%These are listed as follows:
%\begin{itemize}
%	\item JPEG-Compression with quality factor = 90\%
%	\item JPEG-Compression with quality factor = 70\%
%	\item Resizing by a factor of 0.5
%	\item Resizing by a factor of 0.8
%	\item Resizing by a factor of 1.5
%	\item Resizing by a factor of 2.0
%	\item Gamma-Correction using $\gamma$ = 0.8
%	\item Gamma-Correction using $\gamma$ = 1.2
%\end{itemize}
%After performing these manipulations, the resulting number of images in Dataset-I became $8459\times(1+8+1) = 84590$.

%Testing has shown that, the model-specific features of an image are \emph{not} rotation-invariant. The features depend on the orientation of the camera in which the image was captured. However, this information of orientation is not present in the test images. The size of all of the test images is $512 \times 512$, which have been cropped from the center of the original image. 
%To help our networks learn these dependencies on orientation,

\section{Methods}
\subsection{Model Proposal}
\begin{figure}[!h]
	\includegraphics[width = 8cm, height=9cm]{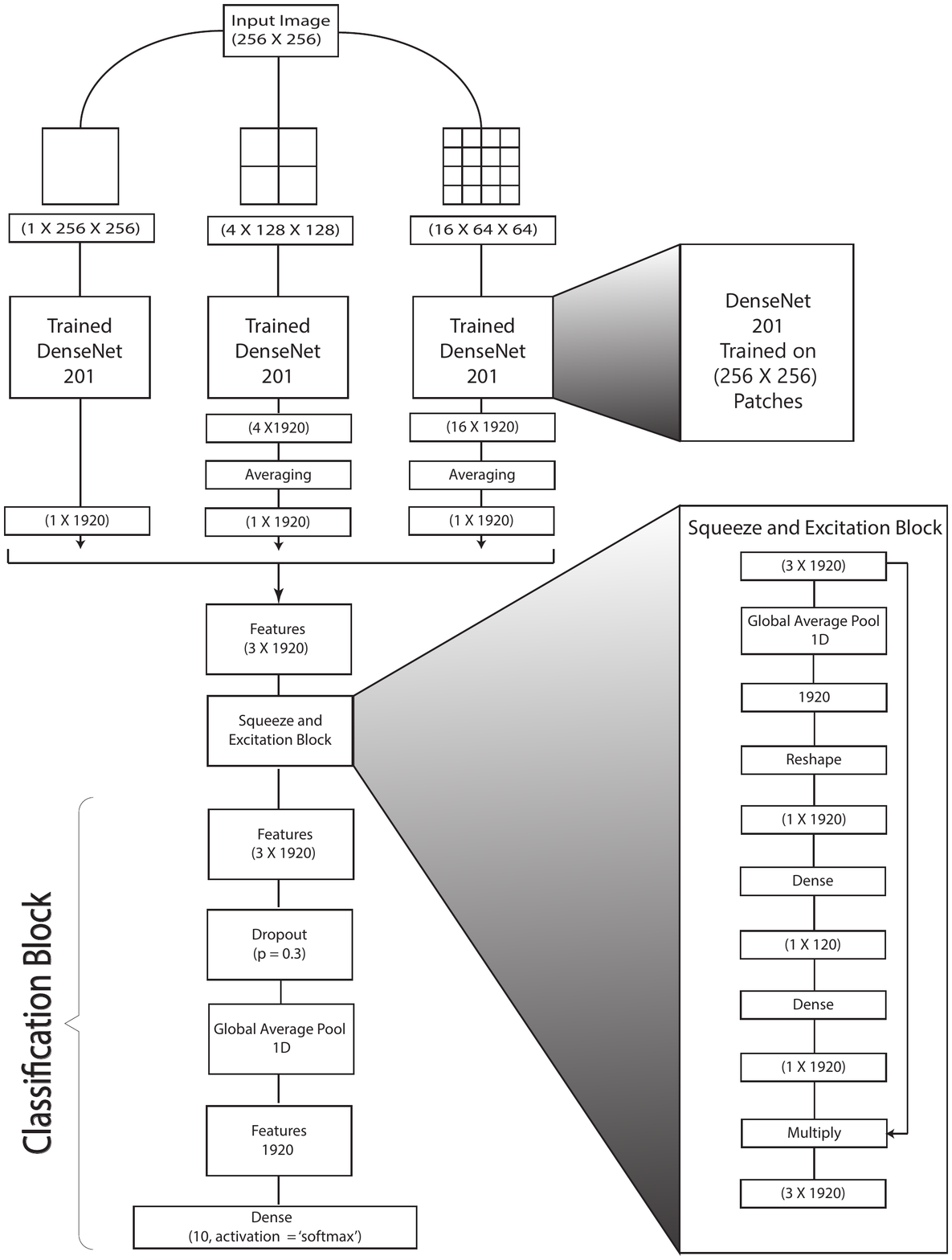}
	\caption{Overview of the network.}
	\label{udayk1}
\end{figure}

The complete structure of our model is shown in Fig. \ref{udayk1}. Different parts of our model are outlined as follows: 
\begin{itemize}  
	\setlength\itemsep{-0.5em}
	\item First, we select patches of size $256\times 256$-- from the generated images based on their quality.
	
	\item After extracting patches, we use them to train Dense Convolutional Networks (DenseNets), specifically the DenseNet-201 architecture, with patches of size $256\times 256$.
	
	\item Next, using the DenseNet-201 trained on $256\times 256$ patches only, we extract features from second to the last layer for the size $256\times 256$ and all non-overlapping patches of size $128\times 128$ and $64\times 64$ from each training image. Thus, at the end, we essentially have 3 feature vectors for 3 different patch size.
	
	\item Then, we concatenate the feature vectors produced by this network and use them to train a secondary network consisting of a Squeeze-and-Excitation (SE) block and a classification block. The output of the SE block is passed to the classification block. 
	
	\item During testing, we 
	%extract all non-overlapping patches of the 3 given sizes from the test image, and, 
	similarly generate feature vectors for each $256\times 256$ patch using the DenseNet-201 trained on $256\times 256$ patches only. These features are concatenated and passed to the secondary network to generate the final prediction for the entire image. 
\end{itemize}

%\begin{figure*}[!t]
%	\centering
%	\includegraphics[width = 13cm, height=10cm]{udayk1.png}
%	\centering
%	\caption{Overview of the network.}
%	\label{udayk1}
%\end{figure*}

\subsection{Training Data Generation}
\subsubsection{Selecting and Extracting Patches}
The first step of the proposed pipeline is to generate patches from the input images---both processed and unprocessed. The idea of extracting patches is motivated by 3 reasons: (i) it results in more data to train our neural network, thus making the training process more generalized; (ii) it enables us to generate multiple predictions for a given test image. Averaging over all of those predictions will ensure a more accurate classification, and (iii) training our network with patches of small size relative to the image prevents our network from learning dominant spacial features of the image. As a result, the network can better learn inherent statistical features related to the source camera model.  The patch sizes that we opt to use is  
%$64\times 64, 128\times 128$ and 
$256\times 256$.

%The patch sizes that we opt to use is  
%%$64\times 64, 128\times 128$ and 
%$256\times 256$. Since, the dimensions of the test images are $512\times 512$, using this size allows us to take non-overlapping patches of the input image. 

But, it is apparent that, all the patches are not suitable for training. In particular, saturated patches are not likely to contain enough statistical information about the used camera model. Therefore, before extracting patches, we determine their quality and only use patches of good quality to train our network. 

We compute the quality value of a patch as outlined in \cite{bondi2017first}. For each patch $\mathcal{P}$ in an image, its quality $Q(\mathcal{P})$ is computed as:
\begin{equation}
\label{eqn:1}
Q(\mathcal{P}) = \frac{1}{3} \sum_{c\, \in [R, G, B]} \left[\alpha\cdot\beta\cdot (\mu_c - \mu_{c}^2)\\
+ (1 - \alpha)\cdot (1 - e^{\gamma \sigma_c} )\right]
\end{equation}
where $\alpha$, $\beta$ and $\gamma$ are empirically set constants (set to 0.7, 4 and $\ln (0.01)$ in our experiments, respectively), whereas $\mu_c$ and $\sigma_c$, $c \in [R, G, B]$ are the mean and standard deviation of the red, green and blue components (normalized  by 255 to the range [0,1]) of patch $\mathcal{P}$, respectively. This quality measure tends to be lower for overly saturated or flat patches, whereas it is higher for textured patches showing some statistical variance. For each image, we select 20 patches of size $256\times 256$ with the highest $Q$ values.

%We use patches of 3 different sizes to make the classification using the secondary network. The spacial dependence of the camera model features becomes very unpredictable if the image is subject to post-processing operations like resizing. In that case, the source-camera features are likely to be scaled up or down. If the discriminant camera-features cannot be identified from a patch of certain size, our model can use patches of other sizes to make a better prediction. Thus, using patches of different sizes provides a certain amount of immunity from these spacial transformations and helps to make a better prediction. 

\subsubsection{Empirical Mode Decomposition}
\begin{figure}[!h]
	\includegraphics[width=8cm, height=5cm]{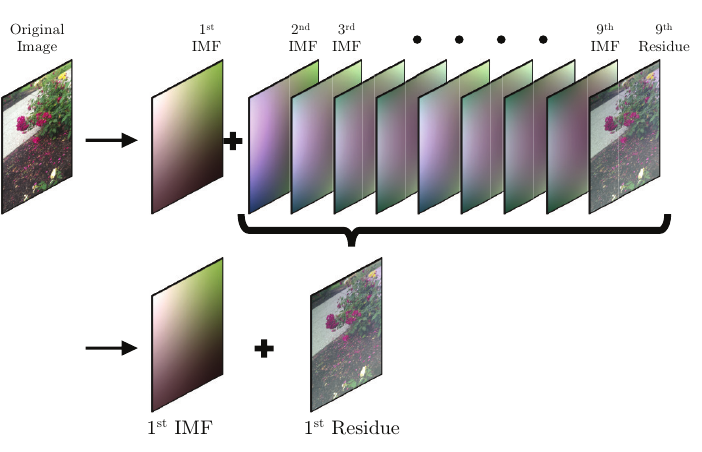}
	\caption{EMD of an Image-- showing the Intrinsic Mode Functions (IMFs) and Residue.}
	\label{udayk2}
\end{figure}

The training data has been augmented further by performing EMD \cite{huang1998empirical}. In EMD, an input signal is decomposed into the so-called Intrinsic Mode Functions (IMFs) and a Residue (see in Fig. \ref{udayk2}). Mathematically, for 2D EMD the deocomposition can be expressed as 
\begin{equation}
\label{eqn:3}
I(m,n) = \sum_{j=1}^{L}\textrm{IMF}_j(m,n) + \textrm{Res}_L(m,n)
\end{equation}
where $\displaystyle I(m,n)$ is the 2D image, $\displaystyle \textrm{IMF}_{j}(m,n)$ is the $\displaystyle j-$th Intrinsic Mode Function, and $\displaystyle \textrm{Res}_{L}(m,n)$ represents the Residue corresponding to $\displaystyle L$ intrinsic modes. \\
In our experiments, we have $\displaystyle m = n = 256$.
The most commonly used algorithm for 2D-EMD is implemented using FastRBF \cite{carr2001reconstruction}. At first, a set of discrete nodes denoted by $\mathbf{X} = \left \{\mathbf{x_i}  \right \}_{i=1}^N \in I(m,n)$ are selected, which are either local minima or local maxima points for $\displaystyle I(m,n)$. Here, $\mathbf{x_i}$ can be described as $(x_i,y_i)$ points on a 2D plane. These coordinates are used as centers for RBF or Radial Basis Functions. 
An RBF or Radial Basis Function \cite{nunes2005texture} is mathematically expressed as
\begin{equation}
\label{eqn:4}
s(\mathbf{x}) = \textrm{p}_m(\mathbf{x}) + \sum_{i=1}^{N}\mathbf{\lambda}_i\mathbf{\phi}(\lVert \mathbf{x} - \mathbf{x_i}\rVert)
\end{equation}
where, $\displaystyle s(\mathbf{x})$ is the Radial Basis Function or RBF, $\displaystyle \textrm{p}_m(\mathbf{x})$ is a low-degree polynomial with degree $\displaystyle m$, $\displaystyle \lambda_i$ are the RBF coefficients, $\displaystyle \phi$ is a real valued function (the spline function is used in our case) and $\displaystyle \mathbf{x}$ denotes variable point $(x,y)$ on 2D space and $\displaystyle \mathbf{x_i}$ are the RBF centers. Here, $\lVert \cdot \rVert$ denotes the Euclidean norm.

%\begin{itemize}
%	\item $\displaystyle s(\mathbf{x})$ is the Radial Basis Function or RBF
%	\item $\displaystyle \textrm{p}_m(\mathbf{x})$ is a low-degree polynomial with degree $\displaystyle m$
%	\item $\displaystyle \lambda_i$ are the RBF coefficients
%	\item $\displaystyle \phi$ is a real valued function (the spline function is used in our case)
%	\item $\displaystyle \mathbf{x}$ denotes variable point $(x,y)$ on 2D space and $\displaystyle \mathbf{x_i}$ are the RBF centers. 
%	Here, $\lVert \cdot \rVert$ denotes the Euclidean norm. 
%\end{itemize}

The algorithm \cite{qiao2008novel} uses FastRBF to interpolate upper and lower envelopes of scattered local maxima and minima from $\displaystyle I(m,n)$. The mean of the envelopes is then subtracted from the image to get the IMF. 

In this work, we have used 2-dimensional EMD to remove the 1st IMF from each channel of the input images separately and retain the residue obtained after the 1st stage decomposition. This residue serves as additional data to train our networks. 
Applied in this manner, EMD essentially works as a denoising scheme by removing random high-frequency noise-components from the image data. Thus, using EMD more than once may prove to be detrimental, as the intrinsic camera-model features embedded in the image may be removed upon successive decompositions.

%We have used Python's PyEMD library to apply the decomposition to all of the $256\times 256$ patches extracted from the SP-Cup Data.\\

\subsection{Architecture}
\subsubsection{Densenet}

%\begin{figure*}[!bht]
%\centering
%\includegraphics[width=6.5in]{udayk3.png}
%%\centering
%\caption{Illustration of Dense-Connections and Transition-Layers implemented in DenseNet.}
%\label{udayk3}
%\end{figure*}

\begin{figure}[!h]
	\includegraphics[width=8cm, height=5cm]{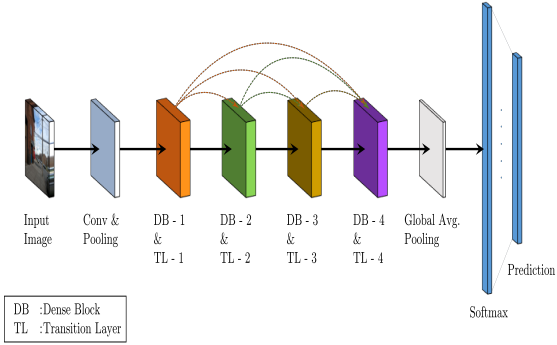}
	\caption{Illustration of Dense-Connections and Transition-Layers implemented in DenseNet.}
	\label{udayk3}
\end{figure}

After extracting patches from the images, we train a deep Convolutional Neural Network (CNN) to perform the task of Source Camera-Model Identification. The CNN model that we opt to use is the Dense Convolutional Network (DenseNet). The details of the DenseNet architecture that we use for Camera-Model Identification is summarized in Table \ref{densearch}.

\begin{table}[!ht]
	\centering
	\caption{Architecture of DenseNet-201}
	\begin{tabular}{|c|c c c c|}
		\hline
		Layers & \multicolumn{4}{|c|}{DenseNet-201}\\
		\hhline{|=|====|}
		Convolution & \multicolumn{4}{|c|}{7 $\times$ 7 conv, stride 2}  \\
		\hline
		Pooling & \multicolumn{4}{|c|}{3 $\times$ 3 max pool, stride 2} \\
		\hline
		Dense Block  & \multirow{2}{*}{\Big[} & 1 $\times$ 1 conv & \multirow{2}{*}{\Big]}  \multirow{2}{*}{$\times$ 6} & \\
		(1) & &  3 $\times$ 3 conv & & \\
		\hline
		Transition Layer & \multicolumn{4}{|c|}{1 $\times$ 1 conv} \\
		\cline{2-5}
		(1)  & \multicolumn{4}{|c|}{3 $\times$ 3 max pool, stride 2} \\
		\hline
		Dense Block  & \multirow{2}{*}{\Big[} & 1 $\times$ 1 conv & \multirow{2}{*}{\Big]}  \multirow{2}{*}{$\times$ 12} & \\
		(2) &  & 3 $\times$ 3 conv & & \\
		\hline
		Transition Layer  & \multicolumn{4}{|c|}{1 $\times$ 1 conv} \\
		\cline{2-5}
		(2) &  \multicolumn{4}{|c|}{2 $\times$ 2 average pool, stride 2}\\
		\hline
		Dense Block  & \multirow{2}{*}{\Big[} & 1 $\times$ 1 conv & \multirow{2}{*}{\Big]}  \multirow{2}{*}{$\times$ 48} & \\
		(3) & & 3 $\times$ 3 conv & & \\
		\hline
		Transition Layer  &\multicolumn{4}{|c|}{1 $\times$ 1 conv} \\
		\cline{2-5}
		(3) & \multicolumn{4}{|c|}{2 $\times$ 2 average pool, stride 2}\\
		\hline
		Dense Block  & \multirow{2}{*}{\Big[} & 1 $\times$ 1 conv & \multirow{2}{*}{\Big]}  \multirow{2}{*}{$\times$ 32} & \\
		(4)  & & 3 $\times$ 3 conv & & \\
		\hhline{|=====|}
		Layer  & \multicolumn{4}{|c|}{Global Average Pooling} \\
		\hhline{|=====|}
		Classification  & \multicolumn{4}{|c|}{Softmax} \\
		\hline
	\end{tabular}
	\label{densearch}
\end{table}

The model that we use is the 201-layer DenseNet introduced in \cite{huang2017densely}. It consists of 4 dense blocks, each with a growth-rate of 32. Transition layers have been used between successive dense blocks. These consist of a convolution layer and a max-pooling layer.  No reduction and dropout layers have been used in the network. The dimensionality of the output feature vector is reduced by using Global Average Pooling, and the features are finally classified by using a Fully-Connected layer with Softmax as the activation function. This layer outputs the probabilities of classification for each class. 

The intuition behind using this architecture is the nature of the classification that we wish to accomplish. The camera-model features inherent in an image are very subtle and minute features of the image \cite{stamm2013information}. Detecting and classifying these features are difficult in itself. But, the task is made even more challenging by the constraints posed for the task. In addition to the model-level features, the image also contains device-level features such as the Photo Response Non Uniformity (PRNU) sensor noise \cite{filler2008using} . To detect the source camera-model effectively, we need to take care that the network does not become dependent on this type of sensor noise. In addition to this, post-processing has also been introduced in the dataset which alters the spacial structure of the model-features in an unpredictable manner.
Therefore, a network that can detect the model-features under all of these constraints needs to be sufficiently deep and have a large number of parameters. But, training such a deep network to detect the subtle model-features proves to be very difficult. The network invariably becomes dependent on the image content or the device specific noise, as all of the minute statistical information is lost when the image is propagated through consecutive layers.

This problem is alleviated in the DenseNet through the use of dense connections. To preserve image information throughout the network, the output of each layer is propagated to all of the layers in front of it. Even if some of the minute features are lost due to some operation, it is regenerated from the output of the previous layers at the input of the subsequent layers through these dense connections (see Fig. \ref{udayk3}). This prevents the gradient-flow from vanishing during training in such a deep network and allows us to extract features which are very difficult - if not impossible to detect using conventional CNN architectures. 

\subsubsection{Squeeze and Excitation Block}
The output after the 4 dense blocks is passed to another module called a "\emph{Squeeze-and-Excitation}" (SE) block. This module has been introduced by Hu, Shen and Sun \cite{hu2017squeeze}. The aim of this module is to improve the representational power of a network by explicitly modelling the interdependencies between the channels of its output. To achieve this, the SE block performs feature recalibration, through which it can learn to use global information to selectively emphasize informative features and suppress less useful ones, without changing the dimensions of the feature vector. 

%\begin{figure*}[!t]
%\centering
%\includegraphics[width=1in]{udayk4.png}
%\caption{\textsf{Illustration of a Squeeze-and-Excitation Block.}}
%\label{udayk4}
%\end{figure*}
\begin{figure}[!h]
	\includegraphics[width=8cm, height=5cm]{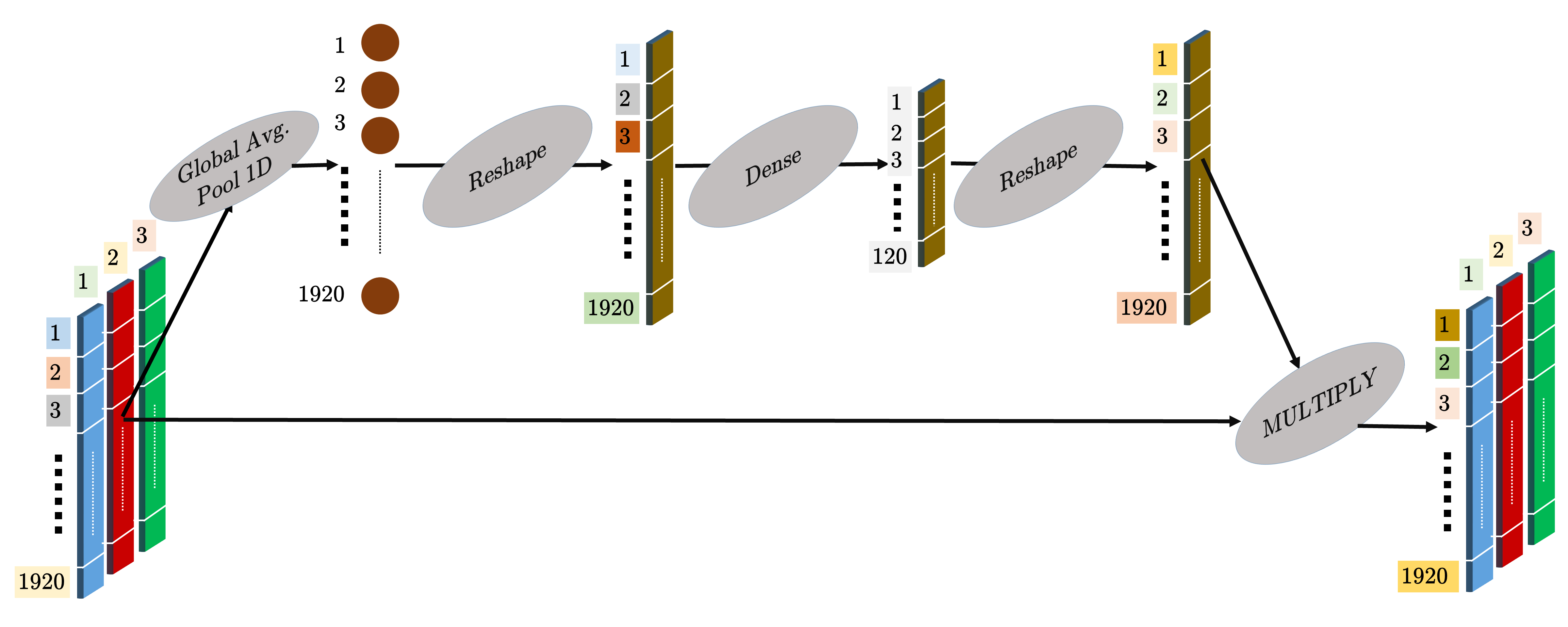}
	\caption{Illustration of a Squeeze-and-Excitation Block.}
	\label{udayk4}
\end{figure}

The internal layers of the SE block and corresponding shapes are given in Figure \ref{udayk4}. We construct the SE block to perform feature recalibration as follows. The input features are first passed through a \emph{squeeze} operation, which aggregates the feature maps across spatial dimensions to produce a channel descriptor. This descriptor embeds the global distribution of channel-wise feature responses, enabling information from the global receptive field of the network to be leveraged by its lower layers. This is followed by an \emph{excitation} operation, in which sample-specific activations, learned for each channel by a self-gating mechanism based on channel dependence, govern the excitation of each channel. The feature maps are then reweighted to generate the output of the SE block which can then be fed directly to the classification layers.

\subsubsection{Classification block}
The modified features, of size $(3\times1920)$, produced at the output of the SE block is then passed through a Dropout layer with a dropout-rate of 30\%. This is followed by a Global Average Pooling operation to reduce the feature vector to a size of $(1\times1920)$. Finally, we pass the pooled feature vector to a Dense Layer with Softmax as the activation function to generate probabilities for the 10 classes which represent the 10 camera models that we need to classify. 

\section{Experiments}
In this section, we discuss the training procedure in detail. Before training, the DenseNet-201 model was initialized using weights pre-trained on the ImageNet database. This ensured a better and faster convergence of the weights during training. 
%All of the experiments regarding training and implementation of the model are performed in hardware environments which included Intel Core-i7 8700K, 3.70 GHz CPUs and Nvidia GeForce GTX 1080 Ti (11 GB Memory) and Nvidia Titan XP (12 GB Memory) GPUs. The necessary codes were written in Python, and the neural network models were implemented by using the Keras API with Tensorflow in the backend.

\subsection{Phase-I}
In Phase-I, we train our model using Dataset-I. We take 20  patches of size $256\times 256$ from each image to train our  network. 
%Taking 20 patches of size $256\times 256$ from each image results in a training dataset of $(93049 \times 20) = 1860980$ segments to train our model. We have also added the EMD data segments in this phase. 
During training, 85\% of the total number of patches are used  for training and the rest are used for validation. We have  used Stochastic Gradient Descent as the Optimizer in our  network with a momentum of 0.9 and initial learning rate of  $10^{-3}$. The learning rate is decreased by a factor of  $10^{-1}$ if the validation loss have not decreased in 2  successive epochs. In this way, when the learning rate is  reduced to $10^{-7}$, training is stopped.

After training the DenseNet, we extract features from the second to last layer for input size of $256\times 256$. The output features are of size $1\times 1920$. We also extract features for all the non-overlapping patches of size $128\times 128$ and $64\times 64$ that the $256 \times 256$ patch contains as visualized in Figure \ref{udayk1}. We receive $4\times 1920$ feature vector from the 4 non-overlapping patches of $128\times 128$ and $16\times 1920$ feature vector from the 16 non-overlapping patches of size $64\times 64$. We reduce both the feature vectors of size $4\times 1920$ and $16\times 1920$ to $1\times 1920$ each by averaging. Lastly, we concatenate the feature vectors for 3 different input sizes to form a resultant feature vector of size $3\times 1920$. It should be noted that the same model is used for extracting features.

We then use the output feature vectors generated by the DenseNet to train our secondary network. These features are passed to the Squeeze-and-Excite and classification network. The classification network outputs a final class vector of size $1\times 10$ which represent the 10 Camera Models that we need to detect (see Figure \ref{udayk1}). 

Although our proposed architecture is a single model trained on $256\times 256$ patches only, for investigation purposes, we have separately trained three separate DenseNet-201 networks for three input sizes, the outcomes of which shall be discussed in the next section.\\

\subsection{Phase-II}
For phase-II, we have used the images of Dataset-II and extracted the best 20 non-overlapping patches of size $256\times256$ depending on the quality we outlined before. 
For training, we have used transfer learning on our previously trained model from Phase-I. We load the weights of the network from Phase-I to initialize the DenseNet and train the network for input size of $256\times 256$. We do not implement our full proposed pipeline in this phase. The output feature of the DenseNet of size $1\times 1920$ for $256\times 256$ input patches have been directly used in classification. The classification block receives the $1\times 1920$ feature vector from DenseNet and is trained for the 27 classes. Both the classification block and the DenseNet runs through the same backpropagation. Hyper-parameters for training have been kept the same as in Phase-I. 
\\
\subsection{Phase-III}
In Phase-III, We have used all images from Phase-I. However, EMD-data has not been included in this case. We sub-divided these data into 4 classes (Unaltered, Resized, JPEG-Compressed and Gamma-Corrected) irrespective of their camera models. 
%The 1st DenseNet obtained from Phase-I is used to extract features from the input image. These features are then used to train a Softmax layer to detect the presence of manipulation in the data.
Similar to phase-II, DenseNet has been initialized using the network from phase-I and the classification block is trained to detect the presence of manipulation in the data. It must be mentioned that, during training, our dataset have been reduced to some extent ($150000\times4 = 600000$) to make the training data evenly distributed among 4 classes. Also, in this case, we have used $128\times128$ sized patches for training due to much higher prediction accuracy compared to other sizes. 

The accuracies obtained from all of these networks during training and testing are included in the result section. 

\section{Results and Discussion}
In this section, we shall discuss our experimental procedure in details. The following subsections will present the outcomes of our experiments.

\subsection{Phase-I}
This is the core result of our work. The test dataset of Phase-I is completely from an unseen device and contains 2640 images of size $512\times512$ with equal numbers of unaltered and manipulated images. We have tested the results generated by our networks in Kaggle. According to the competition rules of IEEE Signal Processing Cup 2018, Kaggle provides a score on the test-results based on the following formula:
\begin{multline*}
\text{Score} =  0.7\times(\text{Accuracy of Unaltered Images}) + \\
0.3\times(\text{Accuracy of Manipulated Images})
\end{multline*}
In this work, whenever we mention overall accuracy, we refer to this score. We can calculate individual accuracies from the above scoring equation by submitting predictions for unaltered or manipulated images separately. The test-accuracies of Phase-I are summarized in Table-\ref{result1}.
\\
\begin{table}[]
	\centering
	\caption{Detection Accuracy of Camera-Models for different Input Sizes}
	\begin{tabular}{|c|c|c|c|}
		\hline
		\multirow{2}{*}{Network}                                                               & \multicolumn{3}{c|}{Accuracy}                                                                                                                                                       \\ \cline{2-4} 
		& \begin{tabular}[c]{@{}c@{}}Unaltered\\ (70\%)\end{tabular} & \begin{tabular}[c]{@{}c@{}}Manipulated\\ (30\%)\end{tabular} & \begin{tabular}[c]{@{}c@{}}Total\\ (100\%)\end{tabular} \\ \hline
		\begin{tabular}[c]{@{}c@{}}DenseNet-201\\ ($64 \times 64$)\end{tabular}   & 67.16\%   & 27.43\%  & 94.59\%                                                  \\ \hline
		\begin{tabular}[c]{@{}c@{}}DenseNet-201\\ ($128 \times 128$)\end{tabular} & 68.33\%  & 28.61\%   & 96.94\%                                                  \\ \hline
		\begin{tabular}[c]{@{}c@{}}DenseNet-201\\ ($256 \times 256$)\end{tabular} & 68.75\%  & 28.82\%   & 97.57\%                                                  \\ \hline
		\begin{tabular}[c]{@{}c@{}}DenseNet-201\\ (Final Layer \\Prediction Average)\end{tabular} & 69.12\%  & 28.84\%  & 97.96\%                                                  \\ \hline
		\begin{tabular}[c]{@{}c@{}}Full\\ Pipeline \end{tabular}  & 69.33\%  & 29.04\%  & 98.37\%                                                 \\ \hline
	\end{tabular}
	\label{result1}
\end{table}

In Table \ref{result1}, we can clearly see the impact of input image size on the test-results. Despite being trained on the same DenseNet-201 architecture, higher accuracy is produced for larger input sizes. It may be the consequence of lower quality of the $64\times 64$ patches. The residual camera-model information left after cropping an image to this size are minimal. This may have caused difficulties for the network to predict accurately for inputs of this size. 
\\
Nonetheless, a better result may be obtained by averaging the predictions of the 3 networks, with separate weights for the classification layer. This illustrates that some of the camera-model features may vary depending on the size of the input. As a result, ensembles over multiple networks trained on different input sizes are likely to have improved performance. 
However, our aim in this work is to maximize the detection-accuracy using a single network. So, we used the weights for the $256\times 256$ input-size in all the networks of our model. We have generated the output features for all the input sizes using this single weight. And we have used the SE network to automatically adjust the weights of these features. This full pipeline achieved an overall accuracy of 98.37\%. 

\begin{figure}[!h]
	\includegraphics[width=8cm, height=5cm]{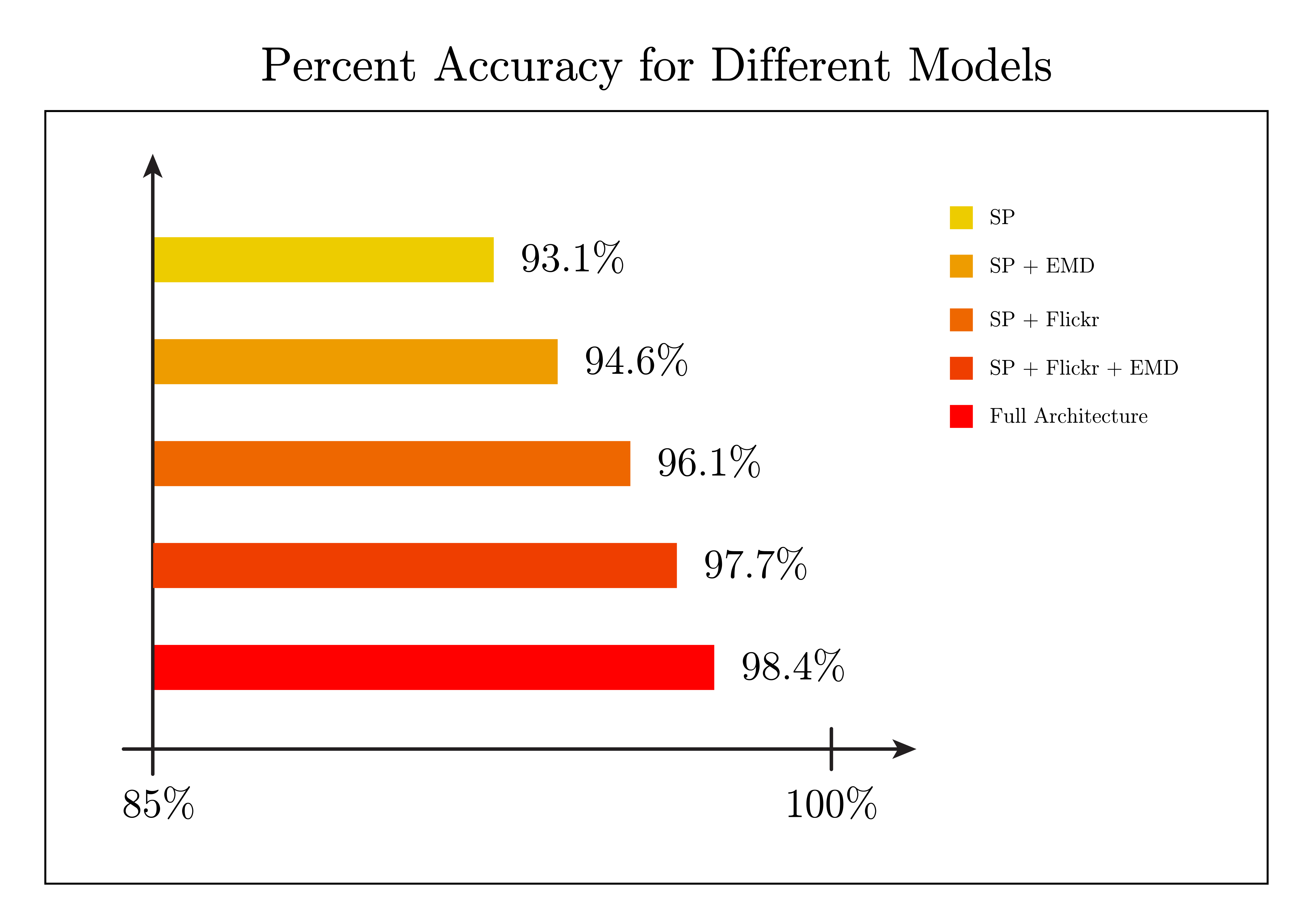}
	\caption{Difference in accuracy with the variation of Dataset. Here we can clearly see the effects of adding EMD augmented images in increasing the accuracy.}
	\label{udayk5}
\end{figure}

Besides, the performance of our network on different parts of the dataset are shown in Figure \ref{udayk5}.

\begin{table}[!hbt]
	\centering
	\caption{Predictions of detected manipulations}
	\begin{tabular}{|l|cccc|}
		\hline
		& \multicolumn{1}{c|}{Unaltered} & \multicolumn{1}{c|}{\begin{tabular}[c]{@{}c@{}}JPEG-\\ Compr.\end{tabular}} & \multicolumn{1}{c|}{\begin{tabular}[c]{@{}l@{}}Gamma-\\ Corrected\end{tabular}} & Resized       \\ \hline
		Unaltered                                                  & \textbf{90.07\%} &3.49\%  & 6.06\%  & 0.38\% \\ \cline{1-1}
		\begin{tabular}[c]{@{}l@{}}JPEG-\\ Compr.\end{tabular}     & 0.15\%  & \textbf{99.85\%}  & 0\%  & 0\% \\ \cline{1-1}
		\begin{tabular}[c]{@{}l@{}}Gamma-\\ Corrected\end{tabular} & 3.18\%  & 0\%  & \textbf{96.75\%}  & 0.07\%  \\ \cline{1-1}
		Resized                                                    & 0\% & 0\% & 0\% & \textbf{100\%} \\ \hline
	\end{tabular}
	\label{result3}
\end{table}

Using patches of size $256\times 256$ only, extracted from the SP Cup Data and their manipulated versions, we achieved a modest accuracy of 93.1\%. However, since these images are collected from only one device for each camera-model, the low result is expected. The network inevitably learns device-level features, which degrades its performance. After adding EMD versions of these images, the accuracy significantly improves to 94.6\%. Since, the 1st IMF of the input image was removed in our EMD versions, this IMF is likely to have some correlations to the device-level features. 
Using the entire Dataset-I during training, which included the data from Flickr, the accuracy is further improved to 96.1\%. This can be mainly attributed to the presence of images captured by different devices in the Flickr Data for each camera-model. Because of the presence of these variations in device-level features, the network could learn the model-specific features more accurately-- thus causing the increase in accuracy. Adding EMD versions of all these images boosted the accuracy to 97.7\%. This strengthens our previous assumption of EMD being an effective augmentation technique in this task. 
To recall, all of these results are achieved by using patches of only one size-- $256\times 256$. Using all of the patches and our full pipeline, we have achieved a final accuracy of 98.37\%. This is our final result on Dataset-I, which is the current state-of-the-art on this dataset using a single network.

\subsection{Phase-II}

For Phase-II, we have tested our network on the images from 27 different camera models of the Dresden Database. Although we have not used our full pipeline for this dataset, an overall accuracy of over 99\% is achieved for 19 camera models by the 1st network of Phase-I.
The camera-models for which accuracy dropped are CANON\_IXUS-55, CANON\_A640, NIKON\_D200, NIKON\_D70, NIKON\_D70S, SONY\_H50, SONY\_T77, SONY\_W170. However, the false detection of images is confined within the models of the manufacturing company. %(Table \ref{result2}). 
It means, this network is able to detect the manufacturing company of the source camera-model with an accuracy of 100\%. Also, we have another very important thing to notice in this case. Though the training dataset is very small ($16961\times20 = 339220$) compared to the dataset of Phase-I, but still, DenseNet-201 is able to detect the camera models very accurately because of the learnt features from Phase-I. This indicates that our network can be fine-tuned to detect further camera models. In case of wrongly detected camera models, such as Nikon\_D200 or Sony\_W170, there is a high chance that these models does have almost similar interpolation method or CFA pattern corresponding to other camera models from the same manufacturer. That is why the test-results show some mismatch for these models. Our findings are in commensurate with the work of Kirchner et al. \cite{kirchner2015forensic} where Nikon D70 and Nikon D70s, have been found out to be the same.

\subsection{Phase-III}

In this phase of experiments, we have tried to identify the 4 types of image-manipulations used on the images of Dataset-I: Unaltered, JPEG Compressed, Gamma Corrected and Resized. In testing, we have used the unaltered images from the test data of size $512\times512$ provided by Kaggle and generated a total of $1320\times9 = 11880$ test images, which include $1320$ unaltered, $2640$ JPEG compressed, $2640$ gamma corrected and $5280$ resized images. Details of the result are given in Table \ref{result3}. We have achieved an overall accuracy of 96.66\% in this task. It is instructive to mention that, these results have been obtained by using only the DenseNet-201 architecture and $128\times128$ patch size.

%In addition, different patch sizes show different identification characteristics as well. In this experiment, larger patch sizes are less efficient in determining manipulation features. Our claim has been justified by the results shown in Table \ref{result4}. As smaller patch sizes are more likely to contain noisy data, it is imperative say that, features of manipulations-- which are basically a part of image noise, can be readily found in smaller image segments compared to larger ones. However, too small patches like $64\times64$ have high probability of having saturated region, from which, extraction of statistical features is very difficult. That is, perhaps, one of the reasons behind this accuracy in $128\times128$ size.

%\begin{table}[t]
%\centering
%\caption{Manipulation Detection Accuracy for different Patch Sizes}
%\begin{tabular}{|c|c|c|c|c|}
%\hline
%Patch Size   & \multicolumn{1}{c|}{Unaltered} & \multicolumn{1}{c|}{\begin{tabular}[c]{@{}c@{}}JPEG-\\ Compr.\end{tabular}} & \multicolumn{1}{c|}{\begin{tabular}[c]{@{}l@{}}Gamma-\\ Corrected\end{tabular}} & Resized       \\ \hline
%($256 \times 256$)  & 33.10\% & 99.81\%  & 97.80\%  & 99.92\% \\ \cline{1-1}
%($128 \times 128$)     & 90.07\%  &99.84\%  & 96.74\%  & 100\% \\ \cline{1-1}
%($64 \times 64$)   & 87.80\% & 100\% & 92.95\% &100\% \\ \hline
%\end{tabular}
%\label{result4}
%\end{table}

The results show that, the features learned by our proposed model, have some sort of orthogonality among them, depending on the type of manipulation present in the image. As a result, these features may be used in other image-forensic tasks outside the premise of our current work. 

\section{Conclusion}
In this paper, we have proposed a DenseNet-oriented pipeline for identifying the source camera-model of an image. We have used DenseNet-201 as well as a Squeeze and Excitation (SE) network for our model architecture and trained our model on the IEEE Forensic Camera-Model Identification Dataset. This pipeline shows an overall accuracy of 98.37\% on the test data provided for the IEEE Signal Processing Cup 2018 Camera Model Identification Challenge. This is the
state-of-the-art result obtained on this dataset using a single network, compared to the winning accuracy of 98.9\% obtained by using an ensemble of 10+ networks. In \cite{kuzin2018camera}, an ensemble of CNNs is used and an accuracy of 98.7\% is reported; this performance has been however obtained using a large amount of scraped data (500Gb of photos) to train their network. A number of Data-Augmentation techniques have been used in our work to extend the dataset, among which EMD is a novel addition to the repertoire of techniques used in Camera-Model Identification. Besides, we have also used transfer learning and evaluated our model on the Dresden Image Database, which showed an accuracy of over 99\% for 19 camera models, where we have been able to detect the manufacturing company of the camera-model with an accuracy of 100\%. However, there is an issue that needs to be addressed regarding the experiment on Dresden image Database. The test images used in this experiment are not from a separate device than the devices used to capture the training images because of the unavailability of multiple devices for all 27 camera models. This may have resulted in higher accuracy than the experiment on SP Cup database. Moreover, the features learned by the DenseNet have also been used to classify the manipulations that have been applied to an image, with an accuracy of 96.66\%. This demonstrates the generalization of our training procedure, for detecting camera-model features across varying datasets and the suitability of using these features in multiple image-forensic tasks.

{\small
\bibliographystyle{ieee_fullname}
\bibliography{rr}

\begin{thebibliography}{10}\itemsep=-1pt

\bibitem{bayar2016deep}
Belhassen Bayar and Matthew~C Stamm.
\newblock A deep learning approach to universal image manipulation detection
  using a new convolutional layer.
\newblock In {\em Proceedings of the 4th ACM Workshop on Information Hiding and
  Multimedia Security}, pages 5--10. ACM, 2016.

\bibitem{bayar2017design}
Belhassen Bayar and Matthew~C Stamm.
\newblock Design principles of convolutional neural networks for multimedia
  forensics.
\newblock {\em Electronic Imaging}, 2017(7):77--86, 2017.

\bibitem{bayram2005source}
Sevinc Bayram, Husrev Sencar, Nasir Memon, and Ismail Avcibas.
\newblock Source camera identification based on cfa interpolation.
\newblock In {\em Image Processing, 2005. ICIP 2005. IEEE International
  Conference on}, volume~3, pages III--69. IEEE, 2005.

\bibitem{bondi2017first}
Luca Bondi, Luca Baroffio, David G{\"u}era, Paolo Bestagini, Edward~J Delp, and
  Stefano Tubaro.
\newblock First steps toward camera model identification with convolutional
  neural networks.
\newblock {\em IEEE Signal Processing Letters}, 24(3):259--263, 2017.

\bibitem{cao2009accurate}
Hong Cao and Alex~C Kot.
\newblock Accurate detection of demosaicing regularity for digital image
  forensics.
\newblock {\em IEEE Transactions on Information Forensics and Security},
  4(4):899--910, 2009.

\bibitem{carr2001reconstruction}
Jonathan~C Carr, Richard~K Beatson, Jon~B Cherrie, Tim~J Mitchell, W~Richard
  Fright, Bruce~C McCallum, and Tim~R Evans.
\newblock Reconstruction and representation of 3d objects with radial basis
  functions.
\newblock In {\em Proceedings of the 28th annual conference on Computer
  graphics and interactive techniques}, pages 67--76. ACM, 2001.

\bibitem{chen2015camera}
Chen Chen and Matthew~C Stamm.
\newblock Camera model identification framework using an ensemble of
  demosaicing features.
\newblock In {\em Information Forensics and Security (WIFS), 2015 IEEE
  International Workshop on}, pages 1--6. IEEE, 2015.

\bibitem{dirik2007source}
A~Emir Dirik, Husrev~T Sencar, and Nasir Memon.
\newblock Source camera identification based on sensor dust characteristics.
\newblock In {\em Signal Processing Applications for Public Security and
  Forensics, 2007. SAFE'07. IEEE Workshop on}, pages 1--6. IEEE, 2007.

\bibitem{filler2008using}
Tom{\'a}s Filler, Jessica Fridrich, and Miroslav Goljan.
\newblock Using sensor pattern noise for camera model identification.
\newblock In {\em Image Processing, 2008. ICIP 2008. 15th IEEE International
  Conference on}, pages 1296--1299. IEEE, 2008.

\bibitem{gloe2012feature}
Thomas Gloe.
\newblock Feature-based forensic camera model identification.
\newblock In {\em Transactions on Data Hiding and Multimedia Security VIII},
  pages 42--62. Springer, 2012.

\bibitem{hu2017squeeze}
Jie Hu, Li Shen, and Gang Sun.
\newblock Squeeze-and-excitation networks.
\newblock {\em arXiv preprint arXiv:1709.01507}, 2017.

\bibitem{huang2017densely}
Gao Huang, Zhuang Liu, Kilian~Q Weinberger, and Laurens van~der Maaten.
\newblock Densely connected convolutional networks.
\newblock In {\em Proceedings of the IEEE conference on computer vision and
  pattern recognition}, volume~1, page~3, 2017.

\bibitem{huang1998empirical}
Norden~E Huang, Zheng Shen, Steven~R Long, Manli~C Wu, Hsing~H Shih, Quanan
  Zheng, Nai-Chyuan Yen, Chi~Chao Tung, and Henry~H Liu.
\newblock The empirical mode decomposition and the hilbert spectrum for
  nonlinear and non-stationary time series analysis.
\newblock In {\em Proceedings of the Royal Society of London A: mathematical,
  physical and engineering sciences}, volume 454, pages 903--995. The Royal
  Society, 1998.

\bibitem{kee2011digital}
Eric Kee, Micah~K Johnson, and Hany Farid.
\newblock Digital image authentication from jpeg headers.
\newblock {\em IEEE transactions on information forensics and security},
  6(3):1066--1075, 2011.

\bibitem{kharrazi2004blind}
Mehdi Kharrazi, Husrev~T Sencar, and Nasir Memon.
\newblock Blind source camera identification.
\newblock In {\em Image Processing, 2004. ICIP'04. 2004 International
  Conference on}, volume~1, pages 709--712. IEEE, 2004.

\bibitem{kirchner2015forensic}
Matthias Kirchner and Thomas Gloe.
\newblock Forensic camera model identification.
\newblock {\em Handbook of Digital Forensics of Multimedia Data and Devices},
  pages 329--374, 2015.

\bibitem{kuzin2018camera}
Artur Kuzin, Artur Fattakhov, Ilya Kibardin, Vladimir~I Iglovikov, and Ruslan
  Dautov.
\newblock Camera model identification using convolutional neural networks.
\newblock In {\em 2018 IEEE International Conference on Big Data (Big Data)},
  pages 3107--3110. IEEE, 2018.

\bibitem{li2008detecting}
Bin Li, Yun~Q Shi, and Jiwu Huang.
\newblock Detecting doubly compressed jpeg images by using mode based first
  digit features.
\newblock In {\em Multimedia Signal Processing, 2008 IEEE 10th Workshop on},
  pages 730--735. IEEE, 2008.

\bibitem{lukavs2006digital}
Jan Luk{\'a}{\v{s}}, Jessica Fridrich, and Miroslav Goljan.
\newblock Digital camera identification from sensor pattern noise.
\newblock {\em IEEE Transactions on Information Forensics and Security},
  1(2):205--214, 2006.

\bibitem{lukas2006digital}
Jan Lukas, Jessica Fridrich, and Miroslav Goljan.
\newblock Digital camera identification from sensor pattern noise.
\newblock {\em IEEE Transactions on Information Forensics and Security},
  1(2):205--214, 2006.

\bibitem{marra2018vulnerability}
Francesco Marra, Diego Gragnaniello, and Luisa Verdoliva.
\newblock On the vulnerability of deep learning to adversarial attacks for
  camera model identification.
\newblock {\em Signal Processing: Image Communication}, 65:240--248, 2018.

\bibitem{marra2017study}
Francesco Marra, Giovanni Poggi, Carlo Sansone, and Luisa Verdoliva.
\newblock A study of co-occurrence based local features for camera model
  identification.
\newblock {\em Multimedia Tools and Applications}, 76(4):4765--4781, 2017.

\bibitem{nunes2005texture}
Jean~Claude Nunes, Steve Guyot, and Eric Del{\'e}chelle.
\newblock Texture analysis based on local analysis of the bidimensional
  empirical mode decomposition.
\newblock {\em Machine Vision and applications}, 16(3):177--188, 2005.

\bibitem{piva2013overview}
Alessandro Piva.
\newblock An overview on image forensics.
\newblock {\em ISRN Signal Processing}, 2013, 2013.

\bibitem{qiao2008novel}
Li-Hong Qiao, Li-Zhong Peng, Wei Guo, and Wei-Tao Yuan.
\newblock A novel image fusion algorithm based on 2d emd and ihs.
\newblock In {\em Machine Learning and Cybernetics, 2008 International
  Conference on}, volume~7, pages 4040--4044. IEEE, 2008.

\bibitem{roy2017camera}
Aniket Roy, Rajat~Subhra Chakraborty, Udaya Sameer, and Ruchira Naskar.
\newblock Camera source identification using discrete cosine transform residue
  features and ensemble classifier.
\newblock In {\em 2017 IEEE Conference on Computer Vision and Pattern
  Recognition Workshops (CVPRW)}, pages 1848--1854. IEEE, 2017.

\bibitem{schmidhuber2015deep}
J{\"u}rgen Schmidhuber.
\newblock Deep learning in neural networks: An overview.
\newblock {\em Neural networks}, 61:85--117, 2015.

\bibitem{stamm2010forensic}
Matthew~C. Stamm and K.J.~Ray Liu.
\newblock Forensic detection of image manipulation using statistical intrinsic
  fingerprints.
\newblock {\em IEEE Transactions on Information Forensics and Security},
  5(3):492--506, 2010.

\bibitem{stamm2013information}
Matthew~C Stamm, Min Wu, and KJ~Ray Liu.
\newblock Information forensics: An overview of the first decade.
\newblock {\em IEEE Access}, 1:167--200, 2013.

\bibitem{swaminathan2007nonintrusive}
Ashwin Swaminathan, Min Wu, and KJ~Ray Liu.
\newblock Nonintrusive component forensics of visual sensors using output
  images.
\newblock {\em IEEE Transactions on Information Forensics and Security},
  2(1):91--106, 2007.

\bibitem{thai2014camera}
Thanh~Hai Thai, Remi Cogranne, and Florent Retraint.
\newblock Camera model identification based on the heteroscedastic noise model.
\newblock {\em IEEE Transactions on Image Processing}, 23(1):250--263, 2014.

\bibitem{tuama2016camera}
Amel Tuama, Fr{\'e}d{\'e}ric Comby, and Marc Chaumont.
\newblock Camera model identification with the use of deep convolutional neural
  networks.
\newblock In {\em Information Forensics and Security (WIFS), 2016 IEEE
  International Workshop on}, pages 1--6. IEEE, 2016.

\bibitem{yang2017source}
Pengpeng Yang, Wei Zhao, Rongrong Ni, and Yao Zhao.
\newblock Source camera identification based on content-adaptive fusion
  network.
\newblock {\em arXiv preprint arXiv:1703.04856}, 2017.

\bibitem{yao2018robust}
Hongwei Yao, Tong Qiao, Ming Xu, and Ning Zheng.
\newblock Robust multi-classifier for camera model identification based on
  convolution neural network.
\newblock {\em IEEE Access}, 6:24973--24982, 2018.

\end{thebibliography}
}

\end{document}